# Discovery of a new magnesium iron boride $Mg_4Fe_{1.1}B_{13.9}$ in the Mg-Fe-B-N system

N.D. Zhigadlo

*CrystMat Company, CH-8037 Zurich, Switzerland*

**Abstract**

A new boron-rich ternary phase, $Mg_4Fe_{1.1}B_{13.9}$, was discovered in the Mg-Fe-B-N quaternary system. This novel phase appears in the form of plate-like shaped crystals formed on the surface of Fe-substituted $MgB_2$ during high-pressure, high-temperature (HPHT) solution growth at 3 GPa and 1960 °C. $Mg_4Fe_{1.1}B_{13.9}$ crystallizes in an orthorhombic structure with space group *Pnam* (No. 62) and lattice parameters $a$ = 10.95(2) Å, $b$ = 7.07(1) Å, and $c$ = 8.72(1) Å. Structural refinement reveals a layered architecture composed of alternating layers of Mg-Fe and boron, with boron atoms forming 5-, 6-, and 7-membered ring motifs. A structural comparison indicates that $Mg_4Fe_{1.1}B_{13.9}$ is closely related to the $Y_2ReB_6$-type borides. This discovery highlights the effectiveness of the HPHT synthesis in accessing new, complex boron-rich phases. As research in the binary Mg-B system approaches saturation, the identification of $Mg_4Fe_{1.1}B_{13.9}$ offers new insights into the formation of phases in the Mg-Fe-B-N system. This paves the way for the guided synthesis of novel borides with unique properties in other multicomponent systems.

*E-mail address:* nzhigadlo@gmail.com. ORCID: 0000-0001-7322-5874

1. ## Introduction

Pressure is a fundamental thermodynamic variable that can be used to manipulate the properties of materials by altering atomic interactions via reducing interatomic distances and reshaping orbital overlaps, allowing the stabilization of exotic bonding motifs and high-coordination frameworks that are inaccessible under ambient conditions [1-5]. Recent advances in both static and dynamic compression methods have led to the discovery and fabrication of numerous functional materials exhibiting remarkable properties [3-5]. These experimental breakthroughs have often been guided and accelerated by first-principles crystal structure prediction methods [6-8]. Notable achievements include the realization of record high-temperature superconductivity in hydrogen sulfide and the synthesis of nanotwinned cubic boron nitride and diamond, which are among the hardest known materials [9,10]. Additionally, high-pressure research has yielded various energy-related and chemically exotic materials [11,12]. However, one persistent challenge is the metastability or decomposition of these materials upon decompression. This issue, along with potential strategies for overcoming it, is addressed in current reviews, which also outline future directions for the field [3].

A key challenge in high-pressure synthesis is associated with the fact that many phases are only stable within a narrow pressure-temperature range and tend to decompose or transform upon decompression to ambient conditions. This issue can often be addressed by rapid quenching, which kinetically traps the high-pressure phase by suppressing atomic mobility. For instance, many copper-based high-temperature superconducting phases, synthesized at 6 GPa and high temperatures, can be quenched to low temperatures to retain its superconducting properties at ambient temperature [13-15]. Alternatively, many high-pressure phases can be stabilized by slow cooling under pressure [16]. In some cases, HPHT conditions are not strictly required for thermodynamic stability, but they significantly facilitate reaction kinetics or enable access to new structural pathways that are otherwise inaccessible under ambient conditions, thus helping to stabilize or discover novel compounds that "freeze" the metastability also after decompression [17]. Interestingly, recently the new concept of so-called of "pressure aging" was also proposed that enables the permanent locking-in of high-pressure structures and their associated enhanced properties in functional materials [18]. Consequently, the implementation of all these impactful approaches introduces a unique dimension to high-pressure materials research.

Borides represent an exciting class of materials for advanced functional applications, as demonstrated by state-of-the-art boride-based superconductors, permanent magnets, catalysts, and topological materials [19]. Earlier comprehensive reviews by Mori [20] and by Albert and Hillebrecht [21] highlight the structural diversity, compositional sensitivity, and rich chemistry of boron-based compounds, where subtle stoichiometric variations yield a wide range of complex boride structures with distinct physical properties. The boride anion ($B^{3-}$) imparts unique electronic and bonding characteristics that are difficult to achieve in other chemical systems, resulting in strong metal-boron bonds and complex extended bonding networks. These features contribute to exceptional structural stability and defect tolerance. Notably, elemental boron itself becomes superconducting under high pressures above ~160 GPa, exhibiting a critical temperature ($T_c$) ranging from 6 to 11 K [22]. To date, hundreds of binary metal borides have been reported, drawing significant attention due to their remarkable properties and potential for diverse technological applications. For example, $ReB_2$, $WB_4$, and $OsB_2$ are among the hardest known materials, exhibiting exceptionally high bulk moduli as a result of strong covalent metal-boron bonding [23]. Beyond mechanical properties, borides are also of great interest for their electronic behavior, particularly superconductivity. $MgB_2$ has attracted significant attention as a prototypical Bardeen-Cooper-Schrieffer (BCS)-type superconductor with the highest superconducting critical temperature ($T_c$) of 39 K among all conventional superconductors [24]. This discovery has stimulated extensive research into crystal growth, elemental substitution, and doping strategies aimed at tuning the superconducting and structural properties of $MgB_2$ and related boride materials [25].

Despite their promising functional properties, ternary and higher-order borides remain relatively unexplored [26]. The scarcity of known compounds in this class, especially those containing Mg, Fe, and B is largely attributed to the challenging requirements for their synthesis. Boron is a refractory element that requires a higher temperature to activate. Methods such as arc melting are commonly used to synthesize transition metal borides. On the other hand, magnesium is a reactive element with a high vapor pressure. The phase diagram shows that Mg and Fe metals cannot form binary compounds, since they are immiscible in both the solid and liquid states [27]. All these facts demonstrate how difficult is to synthesize them under normal conditions. Nonetheless, the inherent structural richness of borides makes them an ideal platform for the discovery of new materials, particularly under extreme synthesis conditions. In recent years, HPHT synthesis has emerged as a powerful strategy for accessing complex boride phases across previously uncharted regions of the chemical space. Under such conditions, boron-based compounds can adopt new metal-to-boron ratios and stabilize unusual structural frameworks, leading to materials with intricate bonding environments and non-traditional stoichiometries [6]. Variations of layered metal-boron structures featuring extended boron ring layers, including the $Y_2ReB_6$-type and related frameworks, have been thoroughly reviewed by Mori [28]. Furthermore, borides often exhibit chemical compatibility with light elements, facilitating the formation of multinary systems with potential applications in high-temperature environments or advanced electronic devices [19]. In this context, the exploration of borides under HPHT conditions offers a compelling pathway toward discovering materials with exceptional structural, mechanical, and electronic properties.

Recently, we employed the HPHT method to investigate Fe substitution in $MgB_2$ single crystals within the Mg-Fe-B-N system [29]. While Fe was successfully incorporated into the $MgB_2$ lattice, the substitution level remained low, at approximately 3%. Unexpectedly, a previously unreported boron-rich ternary phase with composition $Mg_4Fe_{1.1}B_{13.9}$ was observed to form on the surface of some $Mg_{1-x}Fe_xB_2$ crystals. In this work, we report the synthesis, structural characterization, and preliminary insights into the formation conditions of this novel phase. This serendipitous discovery provides valuable information into the high-pressure chemistry of borides and may serve as a guide for the targeted synthesis of crystalline transition-metal borides in other multicomponent systems.

2. **Experimental details**

The starting materials for the single-crystal growth of $Mg_4Fe_{1.1}B_{13.9}$ were high-purity magnesium flakes (99.99%) and powders of iron (99.99%), amorphous boron (99%), and hexagonal boron nitride (99.9%). These materials were weighed in a stoichiometric ratio of 9.5:0.5:12:1 (Mg:Fe:B:hBN) and thoroughly ground in an agate mortar under ambient conditions for about one hour to ensure homogeneity. The resulting mixture was pressed into a pellet of approximately 0.8 cm$^3$ using a hand press and then loaded into a boron nitride (BN) crucible, closed with a BN cup (Fig. 1a). The crucible was placed centrally within a cubic pyrophyllite pressure medium, which served both as a pressure-transmitting and electrically insulating material. The complete high-pressure assembly included a cylindrical graphite resistive sleeve furnace, pyrophyllite insulation tablets, and graphite and stainless-steel disks to provide electrical contacts (Fig. 1b).

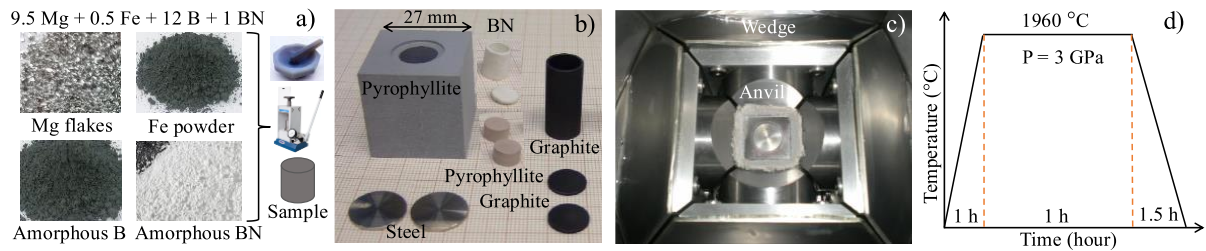

**Fig. 1.** Schematic illustration of the high-pressure, high-temperature (HPHT) synthesis process and sample cell assembly. (a) Starting materials: Mg flakes, Fe powder, amorphous B, and amorphous BN are mixed in a molar ratio of 9.5:0.5:12:1 and cold-pressed into a pellet. (b) HPHT cell assembly comprising a pyrophyllite cube, graphite sleeve heater, pyrophyllite insulating tablets, BN crucible, and stainless steel and graphite disks. (c) Top view on the cubic-anvil cell: the sample, placed in a BN crucible and surrounded by a graphite heater and insulating tablets is embedded in the pyrophyllite cube and compressed between six opposed tungsten carbide (WC) anvils. (d) Temperature-time profile: heating to 1960 °C under 3 GPa over 1 h, held for 1 h, then cooed to room temperature within 1.5 h.

The multi-anvil apparatus consisted of three core components: (1) a 1500-ton uniaxial hydraulic press to generate the required force, (2) a Walker-type module housing both outer and inner anvils arranged symmetrically around the pyrophyllite cube, and (3) the sample assembly itself. Tungsten carbide (WC) anvils compressed the pyrophyllite cube, transmitting quasi-hydrostatic pressure to the BN crucible. As illustrated in Figure 1c, the cubic pressure medium is positioned at the center of symmetry. During pressurization, the pressure medium squeezes out into the spaces between the anvils until the friction between the pressure medium and the anvils balances the pressure generated inside the sample assembly. Thus, the pressure medium must be soft enough to flow at room temperature, but not so soft that it completely squeezes out between the anvils. Another function of the pressure medium is to provide electrical insulation between the furnace, anvils, and sample. This is achieved by in preliminary heating the pyrophyllite cube and the tablets to balance softness and structural integrity, thus ensuring flow during compression while maintaining electrical insulation. Resistive heating was provided by a graphite sleeve furnace, powered by a DC power supply. Temperature control was achieved through the calibrated curve of power input vs. temperature, previously established for this setup. The BN crucible was selected for its high thermal stability, softness, and chemical inertness under HPHT conditions. This method has been extensively used by us to synthesize numerous superconducting [30-34] and magnetic materials [35-37], diamonds [38], cuprate oxides [39,40], pyrochlores [41], polymers [42], and a number of other useful compounds [43-45].

In a high-pressure experiment, the sample is compressed to 3 GPa over 30 minutes, followed by heating up to 1960 °C within 1 hour. This temperature is maintained for 1 hour, after which the system is cooled to room temperature over 1.5 hours (Fig. 1d). Subsequently, a 30-minute decompression is followed before the mechanical disassembly. After that, the pressure medium is carefully broken apart, and the BN crucible extracted. Residual magnesium flux is removed by annealing the as-grown sample at 700 °C under dynamic vacuum. The recovered bulk sample consisted of air-stable, mechanically extractable crystals, displaying plate-like morphology. These included black colored Fe-substituted $MgB_2$ crystals, some with inclusions of new $Mg_4Fe_{1.1}B_{13.9}$ phase, as well as translucent hBN crystals.

Single crystals of $Mg_4Fe_{1.1}B_{13.9}$ were mechanically isolated from the $Mg_{1-x}Fe_xB_2$ matrix under a microscope and characterized using a Bruker SMART CCD diffractometer with Mo Kα radiation (λ = 0.71073 Å) [46]. A total of 1410 frames were collected at room temperature with an exposure time of 30 s per frame and an angular increment Δω = 0.3°. Data processing and structure refinement were carried out using the Bruker software suite and the CrysAlis package for peak indexing, reciprocal layer reconstruction, and intensity extraction. The refinement included mixed occupancy for Mg, Fe, and B atoms and converged to acceptable R-factors.

## 3. Results and discussion

In earlier exploratory studies, we demonstrated that $MgB_2$ crystals can be grown via reactions in the ternary Mg-B-N system [25,47]. Under HPHT conditions, besides elemental Mg, at least four competing phases, such as $MgB_2$, $MgB_4$, hBN, and $MgNB_9$ were observed. Despite the apparent importance of the Mg solvent, the growth process is governed by a complex sequence of phase transformations, which can be summarized as follows [25]:

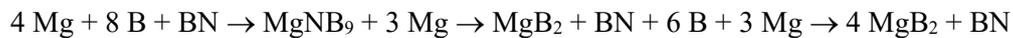

$4\,Mg + 8\,B + BN \rightarrow MgNB_9 + 3\,Mg \rightarrow MgB_2 + BN + 6\,B + 3\,Mg \rightarrow 4\,MgB_2 + BN$

We found that, in the Mg-B-N system, the ternary nitride $MgNB_9$ forms first, and at certain HPHT conditions (see for details refs. 25,47), it decomposes to yield $MgB_2$ crystal seeds. The dissolution of B and $MgB_2$ proceeds further in the excess Mg-rich melt, while hBN single crystals grow simultaneously. In some sense, the Mg-B-N system is unique, since it allows the simultaneous crystal growth of completely different types of materials: a wide-bandgap semiconductor hBN and a superconductor $MgB_2$ [47]. This methodology was also applied to the growth of various substituted $MgB_2$ analogues by partially replacing Mg in the precursor mixture with another element, expanding the ternary Mg-B-N system into more complex quaternary Mg-M-B-N systems (M = Al, C, Mn, Li) [48-52].

Recently, we extended this approach to the Mg-Fe-B-N system in order to grow Fe-substituted $MgB_2$ crystals [29]. While substitution was achieved, the solubility level was low, only up to 3.6 at.% Fe. Interestingly, we observed the formation of a novel boron-rich phase with composition $Mg_4Fe_{1.1}B_{13.9}$ as a surface precipitate on certain $Mg_{1-x}Fe_xB_2$ crystals (Fig. 2).

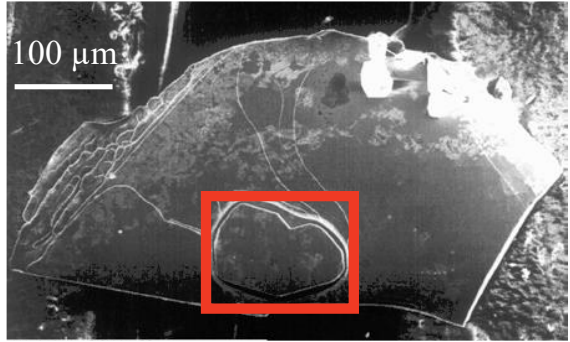

**Fig. 2.** Scanning electron microscopy (SEM) image of a $Mg_{1-x}Fe_xB_2$ crystal grown by the high-pressure method, showing an inclusion of the newly identified $Mg_4Fe_{1.1}B_{13.9}$ phase, highlighted by the red square.

Magnetic susceptibility measurements on these crystals showed a single superconducting transition attributable to $Mg_{1-x}Fe_xB_2$, indicating that $Mg_4Fe_{1.1}B_{13.9}$ is not superconducting. As we show below, the formation of this new phase in the Mg-Fe-B-N multicomponent system under HPHT conditions is inherently complex, driven by a synergy of thermodynamic and kinetic factors and by chemical interactions between the flux and constituent elements. Unlike ambient pressure synthesis, where diffusion is slow and phase equilibria progress slowly and predictably, the HPHT environment promotes high atomic mobility and allows metastable non-equilibrium phases to emerge. In the Mg-Fe-B-N case, the precipitation of the novel $Mg_4Fe_{1.1}B_{13.9}$ phase alongside Fe-substituted $MgB_2$ crystals most likely results from multiple mechanisms.

It is well known that Fe incorporation into $MgB_2$ is inherently challenging. This is due to the different atomic sizes and electronic configurations, as well as to the relatively low reactivity of Fe [29]. At low doping levels, Fe can successfully substitute Mg in the $MgB_2$ lattice, but with increasing Fe content, the substitution kinetics become less favourable, leading to the formation of a distinct phase containing an excess of iron, such as $Mg_4Fe_{1.1}B_{13.9}$. Generally, HPHT conditions increase the atomic mobility, allowing atoms to diffuse more freely, thus increasing the ability of the system to overcome energy barriers and form complex compounds [53]. The crystal structure of the $Mg_4Fe_{1.1}B_{13.9}$ phase might be more adaptable to the high-pressure environment than the simpler $MgB_2$ phase. The $Mg_4Fe_{1.1}B_{13.9}$ phase has a much higher boron content compared to $MgB_2$. Under high pressure, the stability of boron-rich phases increases because the more covalent boron-boron bonds can be stabilized by high pressure. The higher concentration of boron could form a more stable structure that accommodates both magnesium and iron [54,55].

It is worth noting that although nitrogen is not explicitly present in the chemical formula of the newly discovered phase, it may still play a significant role in the reaction environment. Nitrogen can influence phase equilibria in several ways: by altering the chemical behavior of boron (e.g., through the formation of hexagonal BN), acting as a chemical getter, or by modifying the local reaction conditions [25,47]. These effects can shift the equilibrium away from the formation of $Mg_{1-x}Fe_xB_2$, promoting instead the crystallization of more stable ternary borides, such as $Mg_4Fe_{1.1}B_{13.9}$. In the flux-synthesis method, local micro-environments can vary substantially in composition and temperature, potentially favoring the local formation of $Mg_4Fe_{1.1}B_{13.9}$. In addition, strain may play a crucial role, as it is inherently present during the HPHT synthesis. Notably, we observe a preferential nucleation of the $Mg_4Fe_{1.1}B_{13.9}$ phase on the slightly curved surfaces of $Mg_{1-x}Fe_xB_2$ crystals. Curved surfaces typically have higher surface energies compared to flat surfaces, making them energetically favorable sites for nucleation under HPHT conditions. Moreover, these curved regions are often associated with residual strain and a higher density of structural defects, which may locally destabilize the $Mg_{1-x}Fe_xB_2$ lattice and act as a structural trigger for the emergence of a new, energetically favorable ternary boride phase. It is also plausible that local variations in Fe and B concentration occur during the HPHT synthesis, and the curved surface may preferentially accumulate excess Fe and B, especially if diffusivity is enhanced near the crystal-flux interface. This could locally shift the stoichiometry toward conditions that favor the formation of the $Mg_4Fe_{1.1}B_{13.9}$ phase. To conclude, the curved surfaces of $Mg_{1-x}Fe_xB_2$ crystals provide favorable nucleation sites for the $Mg_4Fe_{1.1}B_{13.9}$ phase due to: (i) a higher surface energy and atomic mobility, (ii) localized strain and defect structures, (iii) compositional inhomogeneities, and (iv) enhanced diffusion and crystallization kinetics at the interface.

The new phase, carefully extracted from the $Mg_{1-x}Fe_xB_2$ matrix, was characterized by a CCD diffractometer. The x-ray diffraction patterns taken along the most relevant zone axes are shown in Figure 3.

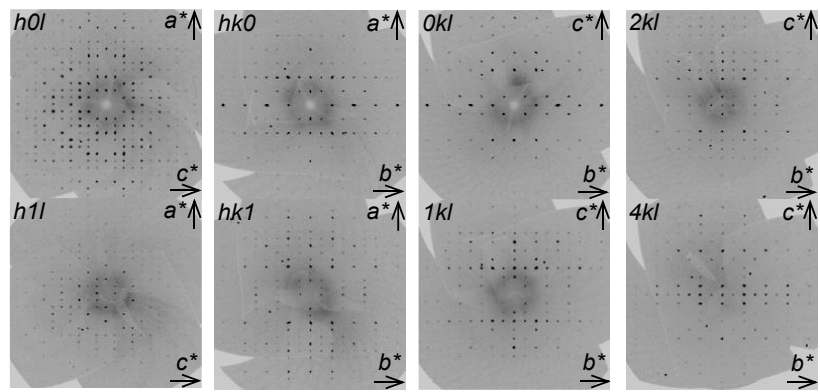

**Fig. 3.** Images of the reconstructed reciprocal space layers of the novel $Mg_4Fe_{1.1}B_{13.9}$ single crystal, obtained using a Bruker SMART CCD diffractometer with MoK$_{\alpha 1}$ radiation ($\lambda$ = 0.71073 Å). Data were collected with an exposure time of 60 s per frame, a detector distance (dd) of 30 mm, and an angular step $\Delta\sigma$ of 0.3°. The intensity of the diffraction spots is represented by their darkness, with darker spots corresponding to higher intensities.

All patterns can be completely indexed with the orthorhombic lattice, and the reflection conditions are compatible with the space group *Pnma* (No. 62). The compound contains four formula units per unit cell ($Z$ = 4), whose refined parameters are $a$ = 10.95(2) Å, $b$ = 7.07(1) Å, and $c$ = 8.72(1) Å. A total of 1410 reflections were measured, and 114 parameters were refined, yielding an excellent refinement quality with $R_{int}$ = 0.030, $R_1$ = 0.021, w$R$ = 0.049, and goodness of fit $S$ = 1.052. The residual electron density peaks were +0.61 and –0.33 e/Å³, indicating a well-resolved and reliable crystal structure (Table 1). During the structural refinement, atomic positions and site occupancies were freely varied, yielding a refined chemical composition of $Mg_4Fe_{1.1}B_{13.9}$. No indication of nitrogen substitution at boron sites was observed. The uncertainty in the refined composition is estimated to be within ±0.01-0.03. The B7-B7 distance in $Mg_4Fe_{1.1}B_{13.9}$ is 1.674 Å, noticeably shorter than the in-plane B-B bonds in $MgB_2$ (1.75-1.78 Å) [49]. This indicates strong bonding interactions and high bond strength within the boron network.

**Table 1.** Single-crystal x-ray diffraction data collection and selected crystallographic details for $Mg_4Fe_{1.1}B_{13.9}$. The crystal structure was refined using CCD data collected from a single crystal of size $0.1 \times 0.07 \times 0.02$ mm$^3$. The compound crystallizes in the orthorhombic space group *Pnma*, which denotes a primitive lattice featuring an *n*-glide plane perpendicular to the *a*-axis, a mirror plane perpendicular to the *b*-axis, and an *a*-glide plane perpendicular to the *c*-axis.

| Composition | $Mg_4Fe_{1.1}B_{13.9}$ |
|---|---|
| Molar mass, gmol$^{-1}$ | 308.91 |
| Temperature, K | 295 |
| Crystal system | orthorhombic |
| Space group | *Pnma* |
| Formula units per cell, Z | 4 |
| Unit cell *a*, Å | 10.95(2) |
| Unit cell *b*, Å | 7.07(1) |
| Unit cell *c*, Å | 8.72(1) |
| Volume $V$, Å$^3$ | 675.07(2) |
| Radiation | MoK$_\alpha$ |
| Wavelength, Å | 0.71073 |
| Diffractometer | Bruker SMART CCD |
| Measured reflections | 1410 |
| No. of parameters refined | 114 |
| $R_{int}$ | 0.030 |
| $R_1$ (observed data) | 0.021 |
| w$R$ (observed data) | 0.049 |
| $S$ (observed data) | 1.052 |
| Largest $\Delta\rho_{max}$; $\Delta\rho_{min}$ (e/Å$^3$) | 0.61; -0.33 |

Interestingly, the boron atoms in $Mg_4Fe_{1.1}B_{13.9}$ form two-dimensional planar nets composed of condensed 5-membered, 6-membered, and 7-membered rings. These boron sheets are sandwiched between layers of metal atoms, with Mg and Fe occupying prismatically coordinated sites. Notably, the larger Mg atoms preferentially occupy the centres of heptagonal prisms, while the smaller Fe atoms reside at the centres of pentagonal prisms. Each boron atom is trigonally prismatically coordinated by surrounding metal atoms, emphasizing the structural complexity and electron-deficient bonding typical of borides (Figs. 4,5)

Overall, the crystal structure of $Mg_4Fe_{1.1}B_{13.9}$ adopts the $Y_2ReB_6$-type structure, which is well known in several rare-earth transition metal boride systems [56,57]. In this structure type, boron atoms frequently form two-dimensional planar networks composed of condensed 5-, 6-, and 7-membered rings [56-60]. Due to its electron deficiency, the metalloid element boron tends to form clusters, stabilizing its structure through multi-centered bonding. This propensity leads to a rich variety of structural motifs, particularly in boron-rich compounds [61]. High-pressure conditions further amplify this structural diversity, enabling the formation of borides with a wide range of compositions and complex boron frameworks. Many of these phases, stabilized by pressure, incorporate intricate B-B bonding networks that would be unstable or energetically unfavorable under ambient conditions. A relevant example is MgNB$_9$, whose structure can be described as an intergrowth of two alternating layers along the *c*-axis: NB$_6$ and MgB$_3$. Within this framework, B$_{12}$ icosahedra in the NB$_6$ layer and B$_6$ octahedra in the MgB$_3$ layer are interconnected, forming a continuous three-dimensional boron framework [62].

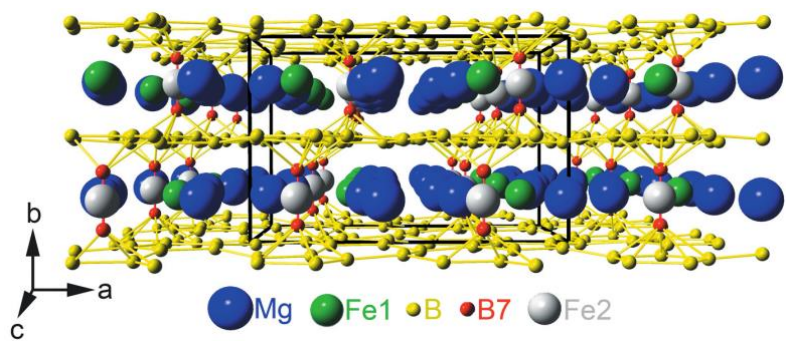

**Fig. 4.** Crystal structure of Mg$_4$Fe$_{1.1}$B$_{13.9}$ viewed along the *c*-axis. This structure features an alternation along the *b*-axis of metal atom layers (comprising Mg and Fe) and boron atom sheets, forming a layered architecture characteristic of this compound: Fe1 (top: drawn in green) position is fully occupied, Fe2: occupation about 10% (top: drawn in grey), either Fe2 or B7 is occupied (B7: occupation about 90%); B7 (top: drawn in red) is connected to the five-membered boron rings (below: drawn in red); B7-B7 = 1.674(3) Å.

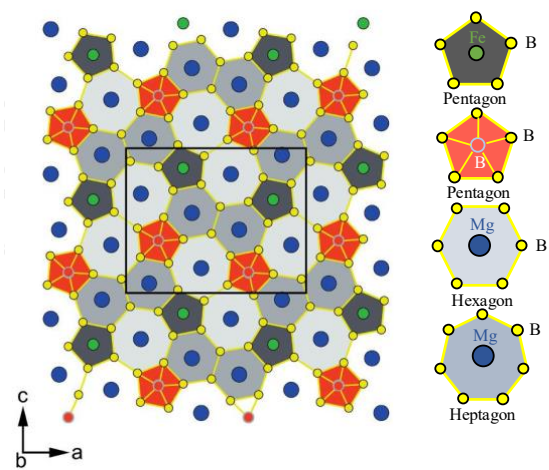

**Fig. 5.** A projection of the crystal structure of Mg$_4$Fe$_{1.1}$B$_{13.9}$ in the *ac* plane, highlighting the network of pentagonal (B$_5$), hexagonal (B$_6$), and heptagonal (B$_7$) boron rings. These structural motifs also shown individually on the right for clarity. The thin lines indicate the unit cell.

The coexistence of B$_5$, B$_6$, and B$_7$ rings in Mg$_4$Fe$_{1.1}$B$_{13.9}$ originates from the electron-deficient nature of boron, its intrinsic tendency to form extended covalent clusters. Unlike electron-rich carbon, which typically favors hexagonal configurations, boron stabilizes diverse polygonal units to compensate for its electron deficiency. Heptagonal B$_7$ rings often arise as natural consequence of strain relaxation and electron deficiency within the boron network, while pentagons and hexagons adjust local curvature and bonding. The combination of these B$_5$, B$_6$, and B$_7$-membered rings effectively reduces internal strain, enhances thermodynamic stability, and results in rigid yet topologically adaptive structures.

These structural features are not unique to $Mg_4Fe_{1.1}B_{13.9}$ and also observed in other boron-based materials, including 2D borophenes, where arrangements of polygonal units depend significantly on substrate interaction and electron doping. Such adaptability of boron frameworks underscores the potential for designing novel borides with tuneable electronic, mechanical, and thermal properties.

Boron electron deficiency renders multiple polygon sizes energetically competitive, and Fe non-stoichiometry in $Mg_4Fe_{1.1}B_{13.9}$ introduces spatially varying charge transfer to the boron framework. Electron-rich regions tend to stabilize larger $B_7$ rings, while electron-poor regions favor smaller $B_5$ rings, with $B_6$ hexagons prevailing in balanced domains. Metal coordination and charge heterogeneity distort the ideal hexagonal network, making the incorporation of non-hexagonal polygons topologically necessary to accommodate local curvature and relieve strain. The preferential formation of $Mg_4Fe_{1.1}B_{13.9}$ on $MgB_2$ crystal surfaces can be explained by the structural compatibility, local chemistry, and HPHT growth kinetics. $MgB_2$ has a planar hexagonal boron sublattice of $B_6$ rings, which provides a suitable template for nucleation. Since the $Mg_4Fe_{1.1}B_{13.9}$ framework retains a significant $B_6$ content, interfacial strain is minimized, thus enabling epitaxial or semi-coherent growth. During synthesis, Fe exhibits a limited solubility in $MgB_2$ but it is readily incorporated into more electron-deficient, distorted boron networks; Fe-rich melt components accumulate at $MgB_2$ surfaces, where their incorporation locally destabilizes the pure $B_6$ network, favoring $B_5/B_7$ ring formation. Kinetically, $MgB_2$ tends to form first as the thermodynamically stable phase in the Mg–B–N system, with Fe-rich fluids becoming enriched near crystal surfaces later in the reaction. Subsequent overgrowth of Fe-containing boron phases thus occurs preferentially at the interface, producing a $MgB_2$ core with a $Mg_4Fe_{1.1}B_{13.9}$ shell. The transformation from a pure $B_6$ network to a mixed $B_5/B_6/B_7$ topology is energetically more favorable at the crystal surfaces, where the boron bonds are unsaturated and flexible, than in the bulk lattice. The $B_6$-membered rings, like in $MgB_2$, contribute to the overall geometric and electronic stability, while $B_5$ and $B_7$ rings emerge from distortions in polyhedral connectivity, especially there where the corner-sharing or edge-sharing deviates from ideal geometry due to local bonding preferences, electron count, or spatial constraints from surrounding metal atoms. In this way, $Mg_4Fe_{1.1}B_{13.9}$ exemplifies how electronic factors, coordination chemistry, and HPHT growth dynamics collectively generate complex boron frameworks with mixed ring sizes.

The discovery of the boron-rich $Mg_4Fe_{1.1}B_{13.9}$ phase expands our understanding of the Mg-Fe-B-N system and suggests that other novel compounds may occur in this system. In the analysis that follows, we used the results of the predictions available from the Open Quantum Materials Database (OQMD) [63] and the Materials Explorer web application, which is a part of the Materials Project next-generation platform [64]. Table 2 shows the compilation of the computationally predicted [63,64,65-67] and experimentally observed [61,62,68-70] ternary compounds in the Mg-Fe-B-N system, along with their crystallographic information. While the actual formation of any given stoichiometry still depends on the synthesis conditions, which cannot be predicted with certainty, the existing dataset reveals clear structural and compositional trends. The Mg–Fe–B–N quaternary system exhibits a remarkable degree of structural and compositional diversity with both known and predicted phases spanning a broad compositional range and adopting crystal structures belonging to nearly all the major crystal systems. Orthorhombic symmetry dominates among the boride and nitridoboride phases, accounting for a substantial number of Mg–Fe–B and Mg–Fe–N compounds (e.g., *Cmcm*, *Pbam*, *Pmmn*, *Pccn*, *Pnma*), reflecting the structural flexibility of boron-rich frameworks in accommodating both Mg and Fe cations. Cubic phases with *Fm-3m* symmetry occur mainly in the Fe-rich borides (e.g., $MgFe_{22}B_6$, $Mg_2Fe_{21}B_6$, $Mg_3Fe_{43}B_{12}$) and in certain nitridoborides (e.g., $Mg(BN_3)_2$), suggesting stabilization by dense metallic sublattices or extened B–N connectivity. Tetragonal and hexagonal structures appear in both borides and nitridoborides, often associated with layered or columnar B and B–N units (e.g., $MgFe_4B_4$, $Mg_9FeB_{20}$, $Fe_5B_2N$), while monoclinic and triclinic structures are observed in several complex nitridoferrates and extended boron frameworks with large formula units (e.g., $Mg_{26}FeB_{54}$, $Mg(Fe_5N_4)_2$, $Mg_9(FeN_4)_2$), possibly due to site-specific ordering or structural frustration.

The dataset includes a number of structurally and compositionally related series such as the $Mg_xFe_yB_{54}$ family (e.g., $Mg_{22}Fe_5B_{54}$ through $Mg_{26}FeB_{54}$), all of which crystallize in the triclinic *P*1 symmetry. This series reflects the incorporation of variable Mg:Fe ratios into large, electron-deficient boron networks, consistent with the known tolerance of boron-rich frameworks for extensive cation substitution and site disorder. Similarly, many Fe-rich borides (e.g., $Mg_3Fe_{20}B_6$, $Mg_5Fe_{41}B_{12}$, $Mg_8Fe_{61}B_{18}$) adopt the *Fm-3m* or closely related cubic symmetries, likely derived from common parent structures by substitution or slight distortions.

**Table 2.** Compilation of experimentally known (noted as E) and predicted by DFT, not yet synthesized (noted as P) ternary compounds in the Mg-Fe-B-N system, along with their crystallographic information. Notes: LP-low pressure phase; HP-high pressure phase; M-monolayer; $T_c$-superconducting transition temperature; data retrieved from the materials project [64] marked with "mp" entry and those retrieved from the Open Quantum Materials Database (OQMD) [63] with "ID" entry.

| Composition | Crystal system | Space group | Reference and note |
|---|---|---|---|
| MgFeB | Orthorhombic | *Cmcm* | [65] P |
| $MgFeB_4$ | Orthorhombic | *Pbam* or *Cmmm* | [66,67] P |
| $MgFe_2B_2$ | Orthorhombic | *Cmmm* | [65] P |
| $MgFe_3B_2$ | Trigonal | *R-3 (R-3m)* | [65] P |
| $MgFe_4B_4$ | Tetragonal | $P4_2/n$ | [65] P |
| $MgFe_7B_3$ | Tetragonal | $I4_1/amd$ | [65] P |
| $MgFe_{12}B_8$ | Orthorhombic | *Cmce* | [65] P |
| $MgFe_{22}B_6$ | Cubic | *Fm-3m* | [65] P |
| $Mg_2FeB_6$ | Orthorhombic | *Pbam* | [65,67] P |
| $Mg_2Fe_7B_7$ | Orthorhombic | *Pccn* | [65] P |
| $Mg_2Fe_{21}B_6$ | Cubic | *Fm-3m* | [65] P |
| $Mg_3FeB_7$ | Orthorhombic | *Cmcm* | [65] P |
| $Mg_3Fe_{20}B_6$ | Cubic | *Fm-3m* | [65] P |
| $Mg_3Fe_{43}B_{12}$ | Cubic (Trigonal) | *Fm-3m (R3m)* | [65] P |
| $Mg_4Fe_{1.1}B_{13.9}$ | Orthorhombic | *Pnam* | [present work] E |
| $Mg_5Fe_{10}B_{24}$ | Orthorhombic | *Pmmn* | [65] P |
| $Mg_5Fe_{18}B_{18}$ | Orthorhombic | *Pccn* | [65] P |
| $Mg_5Fe_{41}B_{12}$ | Cubic (Trigonal) | *Fm-3m (R3m)* | [65] P |
| $Mg_6FeB$ | Orthorhombic | *Amm*2 | [64] P, mp-1022325 |

| Mg$_7$Fe$_{24}$B$_{24}$ | Tetragonal | P-42$_1$c | [65] P |
| Mg$_8$Fe$_{61}$B$_{18}$ | Cubic | Fm-3m (P-3m1) | [65] P |
| Mg$_9$FeB$_{20}$ | Hexagonal | P6/mmm (P-1) | [65] P |
| Mg$_{11}$Fe$_{40}$B$_{40}$ | Tetragonal | P-42$_1$c | [65] P |
| Mg$_{14}$FeB | Orthorhombic | Amm2 | [64] P, mp-1027782 |
| Mg$_{22}$Fe$_5$B$_{54}$ | Triclinic | P1 | [67] P |
| Mg$_{23}$Fe$_4$B$_{54}$ | Triclinic | P1 | [67] P |
| Mg$_{24}$Fe$_3$B$_{54}$ | Triclinic | P1 | [67] P |
| Mg$_{25}$Fe$_2$B$_{54}$ | Triclinic | P1 | [67] P |
| Mg$_{26}$FeB$_{54}$ | Triclinic | P1 | [67] P |
| MgB$_2$N, M | Orthorhombic | Cmmm | [71] P, $T_c$ ~ 11 K |
| Mg(BN$_3$)$_2$ | Cubic | Fm-3m | [64] P, mp-1206547 |
| MgB$_9$N | Trigonal | R-3m | [62] E |
| Mg$_2$BN, M | Tetragonal | I4mm | [71] P, $T_c$ ~ 31 K |
| Mg$_2$BN, M | Monoclinic | Cm | [71] P, $T_c$ ~ 19 K |
| Mg$_2$BN, M | Rhombohedral | R3m | [71] P, $T_c$ ~ 4.5 K |
| Mg$_2$B$_4$N$_2$, M | Hexagonal | P-3m1 | [72] P, $T_c$ ~ 21.8 K |
| α-Mg$_2$B$_4$N$_2$, M | Hexagonal | P-3m1 | [72] P, $T_c$ ~ 7.4 K |
| Mg$_3$BN$_3$ LP | Hexagonal | P6$_3$/mmc | [68] E |
| Mg$_3$BN$_3$ HP | Orthorhombic | Pmmm | [69] E |
| Mg$_3$B$_2$N$_4$ | unknown | unknown | [61] E |
| Mg$_{0.4}$Fe$_{0.6}$N | Cubic | Fm-3m | [70] E |
| MgFeN | Orthorhombic | Pnma | [64] P, mp-1246244 |
| Mg(FeN)$_2$ | Tetragonal | P-4m2 | [64] P, mp-1395522 |
| MgFe$_5$N$_4$ | Orthorhombic | Pmmn | [64] P, mp-1246592 |
| Mg(Fe$_5$N$_4$)$_2$ | Monoclinic | C12/m1 | [64] P, mp-1245863 |
| Mg$_2$FeN$_2$ | Monoclinic | C12/m1 | [64] P, mp-1408987 |
| Mg$_3$FeN$_3$ | Monoclinic | P12$_1$/m1 | [64] P, mp-1245565 |
| Mg$_3$(FeN$_2$)$_2$ | Monoclinic | C12/c1 | [64] P, mp-1246096 |
| Mg$_5$FeN$_4$ | Orthorhombic | Pmmn | [64] P, mp-1246412 |
| Mg$_8$Fe$_3$N$_8$ | Monoclinic | C12/m1 | [64] P, mp-1245495 |
| Mg$_9$(FeN$_4$)$_2$ | Monoclinic | C12/m1 | [64] P, mp-1245778 |
| Fe$_3$BN$_3$ | Tetragonal | P4/mmm | [63] P, ID-1777589 |
| Fe$_3$BN$_5$ | Monoclinic | P2/m | [63] P, ID-1784985 |
| Fe$_5$B$_2$N | Tetragonal | I4/mcm | [63] P, ID-1788277 |
| Fe$_{65}$(BN)$_6$ | Monoclinic | P1m1 | [64] P, mp-1097708 |

Compositionally, a significant portion of the known and predicted compounds are boron-rich: over half the compounds contain more than 50 at. % B, with several, such as Mg$_{26}$FeB$_{54}$ and Mg$_4$Fe$_{1.1}$B$_{13.9}$, exceeding 60 at. % B. On the nitrogen side, 26 compounds incorporate nitrogen, either as discrete nitrides or in extended nitridoborides and nitridoferrates. These often adopt the monoclinic or orthorhombic symmetry and are stabilized by the formation of linear or zigzag N–Fe chains or B–N–Fe linkages.

Trends in stoichiometry further suggest correlation with symmetry type. Compounds with high Fe:B ratios (e.g., MgFe$_{22}$B$_6$, Mg$_3$Fe$_{43}$B$_{12}$) often crystallize in high-symmetry cubic structures, while Mg-rich compositions with large unit cells (e.g., Mg$_{22}$Fe$_5$B$_{54}$ to Mg$_{26}$FeB$_{54}$) favor low-symmetry frameworks. Intermediate Mg–Fe–B compositions (e.g., Mg$_5$Fe$_{18}$B$_{18}$, Mg$_4$Fe$_{1.1}$B$_{13.9}$) tend to display more diverse boron ring topologies and often crystallize in orthorhombic space groups. These observations underscore the structural richness of the Mg–Fe–B–N system and its sensitivity to subtle variations in elemental ratios, electronic structure, and synthesis conditions.

The newly discovered phase $Mg_4Fe_{1.1}B_{13.9}$ expands the known structural diversity of this system, crystallizing in the orthorhombic *Pnam* space group and combines $B_5$, $B_6$, and $B_7$ ring motifs. Its emergence under HPHT conditions highlights the importance of non-ambient thermodynamic constraints in stabilizing kinetically trapped or thermodynamically metastable boron-rich phases. When considered alongside known nitridoborides (e.g., $MgB_9N$, $Mg_3BN_3$, $Mg_3B_2N_4$, $Mg_{0.4}Fe_{0.6}N$), this result reinforces the notion that the Mg–Fe–B–N system remains far from fully explored, with substantial potential for the discovery of additional compounds featuring novel structural motifs, especially under HPHT synthesis conditions.

Interestingly, recent high-throughput density-function theory (DFT) and electron-phonon coupling (EPC) calculations within the Mg–B–N chemical space suggest that several monolayered stoichiometries could host superconductivity with modest $T_c$. In particular, multiple polytypes of $Mg_2BN$ are predicted to be metallic and dynamically stable, with calculated $T_c$ values of ~31 K for *I4mm*-$Mg_2BN$, ~19 K for *Cm*-$Mg_2BN$, and ~4.5 K for *R3m*-$Mg_2BN$ [71], while the $MgB_2N$ (*Cmmm*) [71] and $Mg_2B_4N_2$ (*P-3m1*) [72] phases are predicted to superconduct at ~11 K and ~21.8 K, respectively. While in all cases, the $T_c$ values are below $MgB_2$ it confirms the persistence of σ-bond-mediated pairing in B-N networks. In contrast, no bulk Mg-B-N phases beyond $MgB_2$ were predicted to be superconducting. These results highlight the importance of dimensional reduction: the monolayer geometry enables retention of the boron σ-band superconducting character in the presence of nitrogen, opening new possibilities for engineering 2D superconductivity in the Mg-B-N system.

Collectively, the observed structural diversity suggests significant untapped potential within this quaternary space for discovering new compounds, especially under non-ambient conditions. The coexistence of multiple symmetry types, variable bonding motifs (metallic, covalent, and ionic), and the ability of boron networks to adapt to different cationic environments, all point toward a highly versatile chemical system ripe for further exploration using both experimental and computational approaches.

## 4. Conclusions and outlook

The present study demonstrates the potential of exploratory synthesis in multicomponent systems to yield unexpected yet valuable new materials, particularly in boron-rich chemistries, where even small compositional variations can result in significant structural transformations. While attempting to grow $Mg_{1-x}Fe_xB_2$ crystals from the Mg-Fe-B-N system under HPHT conditions, we serendipitously discovered a novel boron-rich phase, $Mg_4Fe_{1.1}B_{13.9}$. Single-crystal x-ray diffraction reveals that $Mg_4Fe_{1.1}B_{13.9}$ is structurally related to the $Y_2ReB_6$-type borides, crystallizing in the orthorhombic space group *Pnam* (No. 62), with lattice parameters $a = 10.95(2)$ Å, $b = 7.07(1)$ Å, and $c = 8.72(1)$ Å. This ternary boride features a layered crystal structure consisting of alternating metal and boron layers, with boron being arranged in $B_5$, $B_6$, and $B_7$-membered rings. The discovery of $Mg_4Fe_{1.1}B_{13.9}$ demonstrates the effectiveness of HPHT methods for accessing complex, previously unattainable boride phases, especially as research into simpler system, such as binary Mg-B, approaches saturation.

The promising results obtained in the Mg-Fe-B-N system suggest that it warrants further systematic exploration. This quaternary system may host additional intermetallic or boron-rich phases with unique structural features and potentially useful physical properties. Investigating such compounds within this space may enable a deeper understanding of the fundamental interactions between these four elements. Beyond the Mg-Fe-B-N system, many ternary, quaternary, and higher-order systems remain experimentally and computationally unexplored. The HPHT synthesis strategy described here can be further applied to these uncharted chemical spaces, not only to discover new materials, but also to identify general trends over broad compositional spaces.

Finally, the serendipitous discovery of $Mg_4Fe_{1.1}B_{13.9}$ highlights a broad truth in high-pressure research: these environments often induce unexpected chemical reactions and stabilize metastable structures, yielding phases that are inaccessible under ambient conditions. While chance has historically played a key role in such findings, advances in computational prediction and automation can make this process increasingly systematic. Structure prediction algorithms (e.g., USPEX [73], CALYPSO [74], X$_{TAL}$ O$_{PT}$ [75]) AI models trained on pressure-dependent phase diagrams [76], and robotic HPHT synthesis platforms now allow researchers to systematically reproduce and even optimize the conditions that once led to accidental breakthroughs [77]. Together, these advancements are accelerating the discovery of new materials in fields ranging from superconductivity to quantum materials, to energy storage and to extreme-environment technologies.

**Acknowledgments**

The author acknowledges support from the Laboratory of Solid-State Physics ETH Zurich and the Department of Chemistry, Biochemistry and Pharmaceutical Science of the University of Bern. The author gratefully acknowledges G. Schuck for expert assistance with the single-crystal x-ray diffraction measurements and insightful discussion regarding the crystallographic analysis and T. Shiroka for careful reading of the manuscript and helpful suggestions.

**References**

[1] P.F. McMillan, *New materials from high-pressure experiments*, Nat. Mater. 1 (2002) 19-25, https://www.nature.com/articles/nmat716.
[2] V.V. Brazhkin, *High-pressure synthesized materials: Treasures and hints*, High Press. Res. 27 (2007) 333-351, https://doi.org/10.1080/08957950701546956.
[3] L. Zhang, Y. Wang, J. Lv, Y. Ma, *Materials discovery at high pressures*, Nat. Rev. Mat. 2 (2017) 17005, https://doi.org/10.1038/natrevmats.2017.5.
[4] H.-K. Mao, X.-J. Chen, Y. Ding, B. Li, L. Wang, *Solids, liquids, and gases under high pressure*, Rev. Mod. Phys. 90 (2018) 015007, https://doi.org/10.1103/RevModPhys.90.015007.
[5] M. Miao, Y. Sun, E. Zurek, H. Lin, *Chemistry under high pressure*, Nat. Rev. Chem. 4 (2020) 508-527, https://www.nature.com/articles/s41570-020-0213-0.

[6] E. Zurek, W. Grochala, *Predicting crystal structures and properties of matter under extreme conditions via quantum mechanics: the pressure is on*, Phys. Chem. Chem. Phys. 17 (2015) 2917-2934, https://doi.org/10.1039/C4CP04445B.

[7] R.J. Needs, C.J. Pickard, *Perspective: role of structure prediction in materials discovery and design*, APL. Mater. 4 (2016) 053210, https://doi.org/10.1063/1.4949361.

[8] A.R. Oganov, C.J. Pickard, Q. Zhu, R.J. Needs, *Structure prediction drives materials discovery*, Nat. Rev. Matter. 4 (2019) 331-348, https://www.nature.com/articles/s41578-019-0101-8.

[9] A.P. Drozdov, M.I. Eremets, I.A. Troyan, V. Ksenofontov, S.I. Shylin, *Conventional superconductivity at 203 kelvin at high pressures in the sulfur hydride system*, Nature 525 (2015) 73-76, https://www.nature.com/articles/nature14964.

[10] Q. Huang, D. Yu, B. Xu, W. Xu, Y. Ma, Y. Wang, Z. Zhao, B. Wen, J. He, Z. Liu, Y. Tian, *Nanotwinned diamond with unprecedented hardness and stability*, Nature 510 (2014) 250-253, https://www.nature.com/articles/nature13381.

[11] E. Takayama-Muromachi, T. Drezen, M. Isobe, N.D. Zhigadlo, K. Kimoto, Y. Matsui, E. Kita, *New ferromagnets of $Sr_8ARe_3Cu_4O_{24}$ (A = Sr, Ca) with an ordered perovskite structure*, J. Solid State Chem. 175 (2003) 366-371, https://doi.org/10.1016/S0022-4596(03)00334-7.

[12] N.D. Zhigadlo, K. Kimoto, M. Isobe, Y. Matsui, E. Takayama-Muromachi, *High-pressure synthesis, crystal structure and magnetic properties of a new cuprate $(Nd,Ce)_{2+x}CaCu_2O_{6+y}$*, J. Solid State Chem. 170 (2003) 24-29, https://doi.org/10.1016/S0022-4596(02)00011-7.

[13] N.D. Zhigadlo, A.T. Matveev, Y. Ishida, Y. Anan, Y. Matsui, E. Takayama-Muromachi, *Homologous series of high-$T_c$ superconductors $(Cu,C)Sr_2Ca_{n-1}Cu_nO_y$ (n = 2,5) and $(Cu,N,C)Sr_2Ca_{n-1}Cu_nO_y$ (n = 3-6) synthesized under high pressure*, Physica C 307 (1998) 177-188, https://doi.org/10.1016/S0921-4534(98)00527-9.

[14] N.D. Zhigadlo, Y. Anan, T. Asaka, Y. Ishida, Y. Matsui, E. Takayama-Muromachi, *High-pressure synthesis and characterization of a new series of V-based superconductors $(Cu_{0.5}V_{0.5})Sr_2Ca_{n-1}Cu_nO_y$*, Chem. Mater. 11 (1999) 2185-2190, https://doi.org/10.1021/cm990109.

[15] N.D. Zhigadlo, A.T. Matveev, Y. Anan, T. Asaka, K. Kimoto, Y. Matsui, E. Takayama-Muromachi, *High-pressure synthesis and properties of a new oxycarbonate superconductors in the Sr-Ca-Cu-N-C-O system*, Supercond. Sci. Technol. 13 (2000) 1246, https://doi.org/10.1088/0953-2048/13/8/322.

[16] N.D. Zhigadlo, *High-pressure hydrothermal growth and characterization of $Sr_3Os_4O_{14}$ single crystals*, J. Cryst. Growth 623 (2023) 127407, https://doi.org/10.1016/j.jcrysgro.2023.127407.

[17] N.D. Zhigadlo, S. Weyeneth, S. Katrych, P.J.W. Moll, K. Rogacki, S. Bosma, R. Puzniak, J. Karpinski, B. Batlogg, *High-pressure flux growth, structural, and superconducting properties of LnFeAsO (Ln = Pr, Nd, Sm) single crystals*, Phys. Rev. B 86 (2012) 214509, https://doi.org/10.1103/PhysRevB.86.214509.

[18] H. Luo, H. Xuan, D. Wang, Z. Du, Z. Li, K. Bu, S. Guo, Y. Mao, F. Lan, F. Liu, et al., *Pressure aging: An effective process to liberate the power of high-pressure materials research*, PNAS 121 (2024) e2416835121. https://www.pnas.org/doi/suppl/10.1073/pnas.2416835121.

[19] H. Chen, X. Zou, *Intermetallic borides: structures, synthesis and applications in electrocatalysis*, Inorg. Chem. Front. 7 (2020) 2248, https://doi.org/10.1039/D0QI00146E.

[20] T. Mori, "Higher Borides", in: *Handbook on the Physics and Chemistry of Rare-Earths*, vol. 38, eds. K.A. Gschneidner Jr., J.-C. Bünzil, V. Pecharsky (North-Holland, Amsterdam), (2008) 105-173, https://doi.org/10.1016/S0168-1273(07)38003-3.

[21] B. Albert, H. Hillebrecht, *Boron: elementary challenge for experimenters and theoreticians*, Angew. Chem. Int. Ed. 48 (2009) 8640-8668, https://doi.org/10.1002/anie.200903246.

[22] M.I. Eremets, V.V. Struzhkin, H.-K. Mao, R.J. Hemley, *Superconductivity in Boron*, Science 293 (2001) 272-274, https://doi.org/10.1126/science.1062286.

[23] G. Akopov, L.E. Pangilinan, R. Mohammadi, R.B. Kaner, *Perspective: Superhard metal borides: A look forward*, APL Mater. 6 (2018) 070901, https://doi.org/10.1063/1.5040763.

[24] J. Nagamatsu, N. Nakagawa, T. Muranaka, Y. Zenitani, J. Akimitsu, *Superconductivity at 39 K in magnesium diboride*, Nature 410 (2001) 63-64, https://doi.org/10.1038/35065039.

[25] J. Karpinski, N.D. Zhigadlo, S. Katrych, R. Puzniak, K. Rogacki, R. Gonnelli, *Single crystals of $MgB_2$: Synthesis, substitutions and properties*, Physica C 456 (2007) 3-13, https://doi.org/10.1016/j.physc.2007.01.031.

[26] G. Akopov, M.T. Yeung, R.B. Kaner, *Rediscovering the crystal chemistry of borides*, Adv. Mater. 29 (2017) 1604506, https://doi.org/10.1002/adma.201604506.

[27] P. Gao, C. Su, S. Shao, S. Wang, P. Liu, S. Liu, J. Lv, *Iron-magnesium compounds under high pressure*, New J. Chem. 44 (2019) 17403-17407, https://doi.org/10.1039/C9NJ02804H.

[28] T. Mori, *Thermoelectric and magnetic properties of rare earth borides: Boron cluster and layered compounds*, J. Solid State Chem. 275 (2019) 70-82, https://doi.org/10.1016/j.jssc.2019.03.046.

[29] N.D. Zhigadlo, R. Puzniak, *High-pressure high-temperature solution growth, structural, and superconducting properties of Fe-substituted $MgB_2$ single crystals*, J. Cryst. Growth 667 (2025) 128244, https://doi.org/10.1016/j.jcrysgro.2025.128244.

[30] N.D. Zhigadlo, *High pressure crystal growth of the antiperovskite centrosymmetric superconductor $SrPt_3P$*, J. Cryst. Growth 455 (2016) 94-98, https://doi.org/10.1016/j.jcrysgro.2016.10.003.

[31] N.D. Zhigadlo, D. Logvinovich, V.A. Stepanov, R.S. Gonnelli, D. Daghero, *Crystal growth, characterization, and point-contact Andreev-reflection spectroscopy of the noncentrosymmetric superconductor $Mo_3Al_2C$*, Phys. Rev. B 97 (2018) 214518, https://doi.org/10.1103/PhysRevB.97.214518.

[32] N.D. Zhigadlo, *Crystal growth and characterization of the antiperovskite superconductor $MgC_{1-x}Ni_{3-y}$*, J. Cryst. Growth 520 (2019) 56-61, https://doi.org/10.1016/j.jcrysgro.2019.05.021.

[33] P.K. Biswas, S.K. Ghosh, J.Z. Zhao, D.A. Mayoh, N.D. Zhigadlo, X. Xu, C. Baines, A.D. Hillier, G. Balakrishnan, M.R. Lees, *Chiral singlet superconductivity in the weakly correlated metal $LaPt_3P$*, Nat. Commun. 12 (2021) 2504, https://doi.org/10.1038/s41467-021-22807-8.

[34] N.D. Zhigadlo, *High-pressure self-flux growth and characterization of Li-deficient $Li_{0.95}FeAs$ single crystals*, J. Cryst. Growth 649 (2025) 127981, https://doi.org/10.1016/j.jcrysgro.2024.127981.

[35] R. Khasanov, Z. Guguchia, I. Eremin, H. Luetkens, A. Amato, P.K. Biswas, C. Rüegg, M.A. Susner, A.S. Sefat, N.D. Zhigadlo, et al., *Pressure-induced electronic phase separation of magnetism and superconductivity in CrAs*, Sci. Reports 5 (2015) 13788, https://www.nature.com/articles/srep13788.

[36] N.D. Zhigadlo, N. Barbero, T. Shiroka, *Growth of bulk single-crystal MnP helimagnet and its structural and NMR characterization*, J. Alloys Compd. 725 (2017) 1027-1034, https://doi.org/10.1016/j.jallcom.2017.07.247.

[37] N.D. Zhigadlo, *High-pressure growth and characterization of bulk MnAs single crystals*, J. Cryst. Growth 480 (2017) 148-153, https://doi.org/10.1016/j.jcrysgro.2017.10.023.

[38] N.D. Zhigadlo, *Spontaneous growth of diamond from MnNi solvent-catalyst using opposed anvil-type high-pressure apparatus*, J. Cryst. Growth 395 (2014) 1-4, https://doi.org/10.1016/j.jcrysgro.2014.03.002.

[39] N.D. Zhigadlo, J. Karpinski, *High-pressure synthesis and superconductivity of $Ca_{2-x}Na_xCuO_2Cl_2$*, Physica C 460-462 (2007) 372-373, https://doi.org/10.1016/j.physc.2007.03.292.

[40] N.D. Zhigadlo, J. Karpinski, S. Weyeneth, R. Khasanov, S. Katrych, P. Wägli, H. Keller, *Synthesis and bulk properties of oxychloride superconductor $Ca_{2-x}Na_xCuO_2Cl_2$*, J. Phys.: Conf. Ser. 97 (2008) 012121, https://doi.org/10.1088/1742-6596/97/1/012121.

[41] G. Schuck, S.M. Kazakov, K. Rogacki, N.D. Zhigadlo, J. Karpinski, *Crystal growth, structure, and superconducting properties of the β-pyrochlore $KOs_2O_6$*, Phys. Rev. B 73 (2006) 144506, https://doi.org/10.1103/PhysRevB.73.144506.

[42] C. Müller, N.D. Zhigadlo, A. Kumar, M.A. Baklar, J. Karpinski, P.Smith, T. Kreouzis, N. Stingelin, *Enhanced charge-carrier mobility in high-pressure crystallizes poly(3-hexylthiophene)*, Macromolecules 44 (2011) 1221-1225, https://doi.org/10.1021/ma102529f.

[43] P.K. Sahoo, S. Memaran, F.A. Nugera, Y. Xin, T.D. Márquez, Z. Lu, W. Zheng, N.D. Zhigadlo, D. Smirnov, L. Balicas, et al., *Bilayer lateral heterostructures of transition metal dichalcogenides and their optoelectronic response*, ACS Nano 13 (2019) 12372-12384, https://doi.org/10.1021/acsnano.9b04957.

[44] L. Khalil, C. Ernandes, J. Avila, A. Rousseau, P. Dudin, N.D. Zhigadlo, G. Cassabois, B. Gil, F. Oehler, J. Chaste, et al., *High p doped and robust band structure in Mg-doped hexagonal boron nitride*, Nanoscale Adv. 5 (2023) 3225-3232, https://doi.org/10.1039/D2NA00843B.

[45] N.D. Zhigadlo, *Exploring 2D materials by high pressure synthesis: hBN, Mg-hBN, b-P, b-AsP, and GeAs*, J. Cryst. Growth 631 (2024) 127627, https://doi.org/10.1016/j.jcrysgro.2024.127627.

[46] Bruker (2002). SMART (Version 5.62), https://bruker.com.

[47] N.D. Zhigadlo, *Crystal growth of hexagonal boron nitride (hBN) from Mg-B-N solvent system under high pressure*, J. Cryst. Growth 402 (2014) 308-311, https://doi.org/10.1016/j.jcrysgro.2014.06.038.

[48] J. Karpinski, N.D. Zhigadlo, G. Schuck, S.M. Kazakov, B. Batlogg, K. Rogacki, R. Puzniak, J. Jun, E. Müller, P. Wägli, et al., *Al substitution in $MgB_2$ crystals: Influence on superconducting and structural properties*, Phys. Rev. B 71 (2005) 174506, https://journals.aps.org/prb/abstract/10.1103/PhysRevB.71.174506.

[49] S.M. Kazakov, R. Puzniak, K. Rogacki, A.V. Mironov, N.D. Zhigadlo, J. Jun, Ch. Soltman, B. Batlogg, J. Karpinski, *Carbon substitution in $MgB_2$ single crystals: structural and superconducting properties*, Phys. Rev. B 71 (2005) 024533, https://journals.aps.org/prb/abstract/10.1103/PhysRevB.71.024533.

[50] K. Rogacki, B. Batlogg, J. Karpinski, N.D. Zhigadlo, G. Schuck, S.M. Kazakov, P. Wägli, R. Puzniak, A. Wisniewski, F. Carbone, et al., *Strong magnetic pair breaking in Mn substituted $MgB_2$ single crystals*, Phys. Rev. B 73 (2006) 174520, https://journals.aps.org/prb/abstract/10.1103/PhysRevB.73.174520.

[51] J. Karpinski, N.D. Zhigadlo, S. Katrych, K. Rogacki, B. Batlogg, M. Tortello, R. Puzniak, *$MgB_2$ crystals substituted with Li and with Li-C: Structural and superconducting properties*, Phys. Rev. B 77 (2008) 214507. https://doi.org/10.1103/PhysRevB.77.214507.

[52] M. Wälle, J. Koch, D. Tabersky, K. Hametner, N.D. Zhigadlo, S. Katrych, J. Karpinski, D. Günther, *Analyses of lithium-doped and pure magnesium diboride using ultraviolet nano- and femtosecond laser ablation inductively coupled plasma mass spectroscopy*, J. Anal. At. Spectrom. 25 (2010) 193-195. https://doi.org/10.1039/B914547H.

[53] J. Gong, Q. Hou, P. Li, J. Zhang, T. Chen, X. Zhang, *Role of high-pressure high-temperature in atomic diffusion and toughening of 47Zr-45Ti-5Al-3V alloy*, Appl. Mat. Today 45 (2025) 102826, https://doi.org/10.1016/j.apmt.2025.102826.

[54] M.M.E. Ali, J. Chen, I. Harran, B. Sun, X. Cai, H. Wang, L. Tao, Y. Chen, *The pressure-induced chemical structures and properties trend for compressed iron-boride compounds*, J. Phys. Chem. Solids 127 (2019) 238-244, https://doi.org/10.1016/j.jpcs.2018.12.028.

[55] A.N. Kolmogorov, S. Curtarolo, *Theoretical study of metal borides stability*, Phys. Rev. B 74 (2006) 224507, https://doi.org/10.1002/anie.200903246.

[56] Yu.B. Kuz'ma, Yu. Prots, Yu. Grin, *Crystal structure of erbium vanadium tantal boride, Er(V$_{0.77}$Ta$_{0.23}$)VB$_6$*, Z. Kristallogr. NCS 218 (2003) 159-160, https://doi.org/10.1524/ncrs.2003.218.jg.171.

[57] A.M. Alekseeva, A.M. Abakumov, P.S. Chizhov, A. Leithe-Jasper, W. Schnelle, Yu. Prots, J. Hadermann, E.V. Antipov, Yu. Grin, *Ternary magnesium rhodium boride Mg$_2$Rh$_{1-x}$B$_{6+2x}$ with a modified Y$_2$ReB$_6$-type crystal structure*, Inorg. Chem. 46 (2007) 7378-7386, https://doi.org/10.1021/ic7004453.

[58] S. Okada, T. Mori, K. Kudou, T. Shishido, T. Tanaka, *Growth and physical properties of Sc$_2$AlB$_2$ crystals*, J. Phys.: Conf. Ser. 176 (2009) 012008, https://iopscience.iop.org/article/10.1088/1742-6596/176/1/012008/meta.

[59] L.P. Salamakha, O. Sologub, B. Stöger, P.F. Rogl, M. Waas, V. Kapustianyk, E. Bauer, *ScRu$_2$B$_3$ and Sc$_2$RuB$_6$, new borides featuring a 2D infinite boron clustering*, Inorg. Chem. 56 (2017) 10549-10558, https://doi.org/10.1021/acs.inorgchem.7b01512.

[60] T. Mori, T. Shishido, Y. Kawazoe, K. Nakajima, S. Okada, K. Kudou, K. Kiefer, K. Siemensmeyer, *High field magnetization of Tm$_2$AlB$_6$ an AlB$_2$-type analogue compound*, J. Phys.: Conf. Ser. 200 (2010) 012127, https://iopscience.iop.org/article/10.1088/1742-6596/200/1/012127.

[61] P. Rogl, *Materials science of ternary metal boron nitrides*, Intern. J. Inorg. Mater. 3 (2001) 201-209, https://doi.org/10.1016/S1466-6049(01)00009-5.

[62] A. Mironov, S. Kazakov, J. Jun, J. Karpinski, *MgNB$_9$, a new magnesium nitridoboride*, Acta Cryst. C58 (2002) i95-i97, https://doi.org/10.1107/S0108270102009253.

[63] S. Kirklin, J.E. Saal, B. Meredig, A. Thompson, J.W. Doak, M. Aykol, S. Rühl, C. Wolverton, *The Open Quantum Materials Database (OQMD): assessing the accuracy of DFT formation energies*, npj Comput. Mater. **1** (2015) 15010. https://doi.org/10.1038/npjcompumats.2015.10. *Open Quantum Materials Database (OQMD)*, Northwestern University, https://oqmd.org (Accessed on August 9, 2025).

[64] A. Jain, S.P. Ong, G. Hautier, W. Chen, W.D. Richards, S. Dacek, S. Cholia, D. Gunter, D. Skinner, G. Ceder, K.A. Persson, *Commentary: The Materials Project: A materials genome approach to accelerating materials innovation*, APL Mater. **1** (2013) 011002, https://doi.org/10.1063/1.4812323.

[65] Z. Zhang, S. Chen, F. Zheng, V. Antropov, Y. Sun, K.-M. Ho, *Accelerated exploration of empty material compositional space: Mg-Fe-B ternary metal borides*, J. Am. Chem. Soc. 146 (2024) 33179-33192, https://doi.org/10.1021/jacs.4c12648.

[66] Y.Sun, Z. Zhang, A.P. Porter, K. Kovnir, K.-M. Ho, V. Antropov, *Prediction of Van Hove singularity systems in ternary borides*, npj Comp. Mater. 9 (2023) 204, https://doi.org/10.1038/s41524-023-01156-8.

[67] G. Wang, *DFT investigation of hydrogen storage materials*, PhD thesis, University of Missouri, Columbia, Missouri, United States (2016), https://scholarsmine.mst.edu/doctoral_dissertations/2520.

[68] H. Hiraguchi, H. Hashizume, O. Fukunaga, A. Takenaka, M. Sakata, *Structure determination of magnesium boron nitride, Mg$_3$BN$_3$, from x-ray powder diffraction data*, J. Appl. Cryst. 24 (1991) 286-292, https://doi.org/10.1107/S0021889891001334.

[69] H. Hiraguchi, H. Hashizume, S. Sasaki, S. Nakano, O. Fukunaga, *Structure of a high-pressure polymorph of Mg$_3$BN$_3$ determined from x-ray powder data*, Acta Cryst. B49 (1993) 478-483, https://doi.org/10.1107/S0108768192013533.

[70] G. Serghiou, G. Ji, N. Odling, H.J. Reichmann, J.P. Morniroli, R. Boehler, D.J. Frost, J.P. Wright, B. Wunder, *Creating reactivity with unstable endmembers using pressure and temperature: synthesis of bulk cubic Mg$_{0.4}$Fe$_{0.6}$N*, Angew. Chem. Int. Ed. 54 (2015) 15109, https://doi.org/10.1002/anie.201506257.

[71] J. Jiang, Y. Xue, L. Zha, S. Yao, B. Wang, W. Hu, L. Peng, T. Shi, J. Chen, X. Liu, J. Lin, *Machine learning and first-principles calculations for the prediction and analysis of superconductivity in Mg-B-N systems*, J. Mater. Chem. C 13 (2025) 9799-9808, https://doi.org/10.1039/D5TC00708A.

[72] D. Wines, K. Choudhary, A.J. Biacchi, K.F. Garrity, F. Tavazza, *High-throughput DFT-based discovery of next generation two-dimensional (2D) superconductors*, Nano Lett. 23 (2023) 969-978, https://doi.org/10.1021/acs.nanolett.2c04420.

[73] C.W. Glass, A.R. Oganov, N.Hansen, *USPEX-Evolutionary crystal structure prediction*, Comp. Phys. Comm. 175 (2006) 713-720, https://doi.org/10.1016/j.cpc.2006.07.020.

[74] Y. Wang, J. Lv, L. Zhu, Y. Ma, *CALYPSO: A method for crystal prediction*, Comp. Phys. Comm. 183 (2012) 2063-2070, https://doi.org/10.1016/j.cpc.2012.05.008.

[75] D.C. Lonie, E. Zurek, X$_{TAL}$ O$_{PT}$: *An open-source evolutionary algorithm for crystal structure prediction*, Comp. Phys. Comm. 182 (2011) 372-387, https://doi.org/10.1016/j.cpc.2010.07.048.

[76] B. Maruyama, J. Hattrick-Simpers, W. Musinski, L. Graham-Brady, K. Li, J. Hollenbach, A. Singh, M.L. Taheri, *Artificial intelligence for materials research at extremes*, MRS Bulletin 47 (2022) 1154-1164, https://doi.org/10.1557/s43577-022-00466-4.

[77] N.J. Szymanski, B. Rendy, Y. Fei, R.E. Kumar, T. He, D. Milsted, M.J. McDermott, M. Gallant, E.D. Cubuk, A. Merchant, et al., *An autonomous laboratory for the accelerated synthesis of novel materials*, Nature 624 (2023) 86-91, https://doi.org/10.1038/s41586-023-06734-w.